# Detecting giant electron-hole asymmetry in graphene monolayer generated by strain and charged-defect scattering via Landau level spectroscopy

Ke-Ke Bai[1,2], Yi-Cong Wei[1,2], Jia-Bin Qiao[1,2], Si-Yu Li[1,2], Long-Jing Yin[1,2], Wei Yan[1,2], Jia-Cai Nie[1], and Lin He[1,2,*]

[1] Department of Physics, Beijing Normal University, Beijing, 100875, People's Republic of China
[2] The Center of Advanced Quantum Studies, Beijing Normal University, Beijing, 100875, People's Republic of China
* Email: helin@bnu.edu.cn

The electron-hole symmetry in graphene monolayer, which is analogous to the inherent symmetric structure between electrons and positrons of the Universe, plays a crucial role in the chirality and chiral tunnelling of massless Dirac fermions. Here we demonstrate that both strain and charged-defect scattering could break this symmetry dramatically in graphene monolayer. In our experiment, the Fermi velocities of electrons $v_F^e$ and holes $v_F^h$ are measured directly through Landau level spectroscopy. In strained graphene with lattice deformation and curvature, the $v_F^e$ and $v_F^h$ are measured as $(1.21 \pm 0.03) \times 10^6$ m/s and $(1.02 \pm 0.03) \times 10^6$ m/s, respectively. This giant asymmetry originates from enhanced next-nearest-neighbor hopping in the strained region. Around positively charged-defect, we observe opposite electron-hole asymmetry, and the $v_F^e$ and $v_F^h$ are measured to be $(0.86 \pm 0.02) \times 10^6$ m/s and $(1.14 \pm 0.03) \times 10^6$ m/s, respectively. Such a large asymmetry is attributed to the fact that the massless Dirac fermions in graphene monolayer are scattered more strongly when they are attracted to the charged-defect than when they are repelled from it.

Charge carriers in graphene monolayer exhibit light-like dispersion and, usually, the electron and hole are symmetric[1,2]. These features of the massless Dirac fermions result in many unique electronic properties in graphene, for example, the chiral tunnelling[3]. Breaking the symmetry of the positively and negatively charged fermions could destroy the chirality of quasiparticles and the chiral tunnelling in graphene systems[3,4]. Transport measurements in graphene in the presence of charged impurities reveal differences in the mobility seen for electrons versus holes[5-7]. Recently, scanning tunneling microscopy (STM) and spectroscopy (STS) measurements demonstrate explicitly that local density of states (LDOS) of electrons and holes in graphene monolayer have quite different behaviors around a Coulomb potential[8-10], indicating the breaking of electron-hole symmetry by charged-impurity scattering[11-16].

In this paper, we address the issue of electron-hole asymmetry in graphene monolayer through Landau level (LL) spectroscopy and we conduct a systematic study of the electron-hole asymmetry both in strained graphene region and around charged-defect. According to the Landau quantization, we could obtain the Fermi velocities of electron-like and hole-like Dirac fermions in graphene monolayer separately[17-19]. This provides unique opportunity to study the electron-hole asymmetry quantitatively.

The graphene monolayer was grown on Rh foil via a traditional ambient pressure chemical vapour deposition method (see Supplementary Information A.1 for details) and we demonstrated previously that the graphene sheet on Rh foil behaves as pristine graphene monolayer[20,21]. Such a decoupling behavior between the graphene sheet and the supporting substrate is further confirmed by carrying out STS measurements under high magnetic fields, as shown in Figure 1. STS spectra recorded in the graphene sheet with different magnetic fields show Landau quantization of massless Dirac fermions, as shown in Fig. 1c. The observed LL energies $E_n$ depend on the square-root of both level index $n$ and magnetic field $B$[17-19]

$$E_n = \text{sgn}(n)\sqrt{2e\hbar v_F^2 |n| B} + E_0, \qquad n = ...-2, -1, 0, 1, 2... \qquad (1)$$

Here $E_0$ is the energy of Dirac point, $e$ is the electron charge, $\hbar$ is the Planck's constant, $v_F$ is the Fermi velocity, and $n > 0$ corresponds to empty-state (holes) and $n < 0$ to filled-state (electrons). The linear fit of the experimental data to Eq. (1), as shown in Fig. 1d, yields a Fermi velocity of $v_F = (1.14 \pm 0.02) \times 10^6$ m/s for both the electrons and holes. There is a negligible electron-hole asymmetry in the graphene sheet on Rh foil, which agrees very well with the results reported in pristine graphene in previous studies[17-19].

The Hamiltonian of graphene monolayer in tight-binding model reads

$$H = -t \sum_{\langle i,j \rangle \sigma} \left( a_{\sigma,i}^+ b_{\sigma,j} + \text{H.c.} \right) - t' \sum_{\langle\langle i,j \rangle\rangle \sigma} \left( a_{\sigma,i}^+ a_{\sigma,j} + b_{\sigma,i}^+ b_{\sigma,j} + \text{H.c.} \right), \qquad (2)$$

where the operators $a_{\sigma,i}^+$ ($a_{\sigma,i}$) create (annihilate) an electron with spin $\sigma$ at site $i$, $t \sim 3$ eV is the nearest-neighbor hopping integral, and $t'$ is the next-nearest-neighbor hopping energy[1,2]. The observed electron-hole symmetry, as shown in Fig. 1, implies that the next-nearest-neighbor hopping $t'$ in the pristine graphene sheet is negligibly small[1,2]. According to the Hamiltonian (2), the simplest method to break the electron-hole symmetry in graphene monolayer is to introduce a nonzero $t'$, which is facile to be realized in strained graphene[21,22]. Previously, asymmetry of local DOS of electrons and holes has been observed in strained graphene[22]. However, a quantitative measurement of the Fermi velocities of electrons and holes has not been reported so far.

Figure 2a and 2b show representative STM images of a strained graphene region on Rh foil. Similar strained structures are easy to be observed for graphene grown on metallic substrates due to mismatch of thermal expansion coefficients between graphene and the supporting substrates[23,24]. The nanoscale ripples, as shown in Fig. 2a and Fig. 2b, are expected to induce a large electron-hole asymmetry. STS spectra recorded in the strained ripples under magnetic fields exhibit Landau quantization of massless Dirac fermions, as shown in Fig. 2c (similar Landau quantization is also observed in other strained nanoripples, as shown in Fig. S1). According to Eq. (1), we could obtain $v_F^e$ and $v_F^h$ separately. The Fermi velocities of electrons $v_F^e$ and holes $v_F^h$ are quite different, as shown in Fig. 2d, and they are measured to be $(1.21 \pm 0.03) \times 10^6$ m/s and $(1.02 \pm 0.03) \times 10^6$ m/s respectively. Additionally, the measured $v_F^e$ and $v_F^h$ are almost independent of positions in the ripple, as shown in Fig. S1. Such a large electron-hole asymmetry is attributed to the enhanced next-nearest-neighbor hopping by lattice deformation and curvature in the strained region[22,25]. By introducing a nonzero $t'$ in graphene, the Fermi velocity of the filled-state (electrons) increases, whereas the Fermi velocity of empty-state (holes) decreases, as shown in Fig. 2d and Fig. 2f. The theoretical result agrees with our experimental result quite well. In the experiment, the $v_F^e$ increases from $(1.14 \pm 0.02) \times 10^6$ m/s to $(1.21 \pm$ 0.03) × $10^6$ m/s, whereas the $v_F^h$ decreases from (1.14 ± 0.02) × $10^6$ m/s to (1.02 ± 0.03) × $10^6$ m/s. A finite value of $t' = 0.16t$ describes well the observed electron-hole asymmetry in the ripple shown in Fig. 2b. Here we should point out that the observed electron-hole asymmetry differs over different strained graphene ripples, as shown in Fig. S2, and the estimated $t'$ ranges from about $0.02t$ to about $0.2t$ in our experiment.

Besides in strained graphene, we will demonstrate subsequently that $v_F^e$ and $v_F^h$ in graphene monolayer could also be quite different around the charged-defect. Figure 3a shows a representative STM image of a graphene sheet with high-density atomic defects. A $\sqrt{3}\times\sqrt{3}R30°$ interference pattern of carbocyclic rings can be observed around the defects, as shown in Fig. 3b. This interference pattern is attributed to the elastic scattering process between the two adjacent Dirac cones at $K$ and $K'$ induced by the atomic-scale defects[26,27]. Our STS spectra, as shown in Fig. 3c, show the existence of a resonance peak above the Dirac point associated with the defect. Similar interference pattern and resonance peak are observed in other atomic-scale defects (see Supplementary Information Fig. S3 for more experimental results), and such features have also been observed previously in graphene with point defects grown on metallic substrates[27-29]. An unexpected result is that the resonance peak of the defects shifts to higher energy with increasing the magnetic fields, as shown in Fig. 3d. The relative shift of the energy position of the peak $E_p$ with respect to the charge neutrality point $E_0$ as a function of the magnetic fields is summarized in Fig. 4b. Such a feature, which has not been explored before, indicates that the value of $E_p$ not only depends on the structure of the defect[27-29], but also depends on the magnetic fields.

Figure 4a shows typical spectra recorded under different magnetic fields around the defects. Obviously, the observed Landau quantization exhibits large asymmetry between the empty-state above the Dirac point and the filled-state below the Dirac point, as shown in Fig. 4c. According to Eq. (1), we obtain the Fermi velocities of electrons and holes. Around the defects (within 10-15 nm), the electron-hole asymmetry is almost independent of the recorded positions, as shown in Fig. 4d. Comparing the structure of the graphene sheet in Fig. 3a with that of the pristine graphene monolayer[17-19], the main differences are the existence of nanoscale rippling and the atomic defects. These nanoscale ripples usually with $h^2/(la) < 0.0254$ cannot induce such a large electron-hole asymmetry (here $h$ is the amplitude, $l$ is the width of the ripple, and $a = 0.142$ nm)[22], and more importantly, the nanoscale ripple with larger $h^2/(la)$, as shown in Fig. 2 for example, exhibit opposite electron-hole asymmetry as that observed around the defects (Fig. 4). Therefore, we attribute the large electron-hole asymmetry shown in Fig. 4 mainly to the charged-defect scattering[11-16]. The electron-like Dirac fermions and hole-like Dirac fermions in graphene monolayer are expected to respond differently to a Coulomb potential: they are scattered more strongly when they are attracted to the charged-defect than when they are repelled from it (Fig. 4e)[11-16]. According to the observed electron-hole asymmetry, these defects shown in Fig. 3a are of positive charge, which is consistent with the fact that the resonance peak of the charged-defect is located above the charge neutrality point.

In summary, the STM and STS measurements presented here demonstrate that strained regions and charged-defect generate giant electron-hole asymmetry in graphene monolayer. Through Landau level spectroscopy, we direct measure the differences in the Fermi velocities of electrons and holes for the first time. The observed giant electron-hole asymmetry in strained graphene regions and around the charged-defect (or charged-adatom) may account for the different mobility of electrons and holes observed in graphene transport measurements[5-7].

**Acknowledgements**

We thank Zhongfan Liu and Mengxi Liu for the help in the synthesis of the sample.

This work was supported by the National Basic Research Program of China (Grants Nos. 2014CB920903, 2013CBA01603, 2013CB921701), the National Natural Science Foundation of China (Grant Nos. 11422430, 11374035, 11474022, 51172029, 91121012), the program for New Century Excellent Talents in University of the Ministry of Education of China (Grant No. NCET-13-0054), Beijing Higher Education Young Elite Teacher Project (Grant No. YETP0238).

**Author contributions**

L.H. conceived and provided advice on the experiment, analysis, and theoretical calculation. K.K.B. performed the STM experiments. K.K.B., Y.C.W., J.B.Q., S.Y.L., L.J.Y., and W.Y. analyzed the data and performed the theoretical calculations. L.H. wrote the paper. All authors participated in the data discussion.

**Competing financial interests:** The authors declare no competing financial interests.

**Figure Legends**

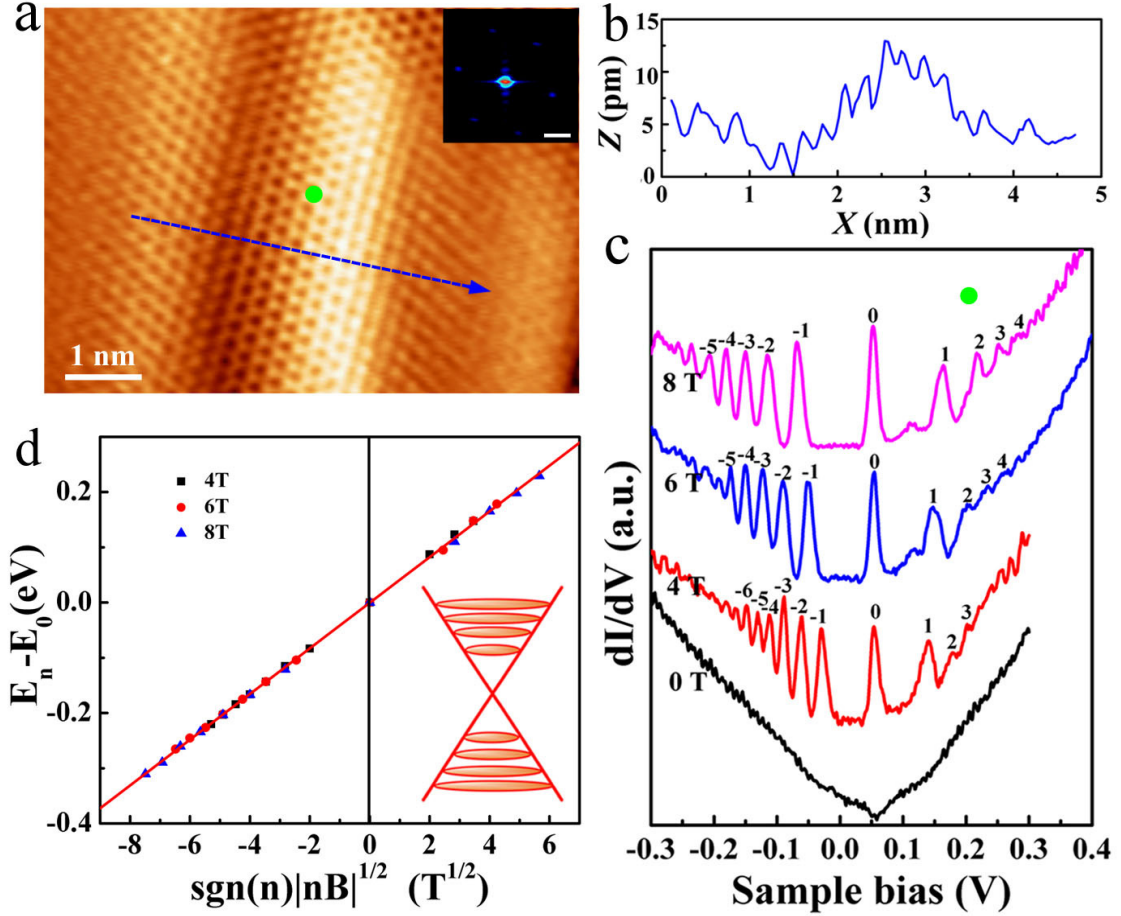

**Figure 1 | STM image and STS spectra of a graphene monolayer on Rh foil. a**, An atomic- resolution STM image of a graphene sheet on Rh foils ($V_{sample}$ = 0.57 V and $I$ = 69 pA). The inset is the Fast Fourier transform showing the reciprocal-lattice of graphene. The scale bar is 15 Gm$^{-1}$. **b**, The height profile along the dashed blue line in **a** indicating ultrafine rippling of the graphene sheet. **c**, STS spectra taken under different magnetic fields at green solid circle marked in panel **a**. For clarity, the curves are offset in Y-axis and LL indices of massless Dirac fermions are marked. **d**, LL peak energies of the spectra show square-root dependence on both the LL indices and magnetic fields, i.e., sgn($n$)|$nB$|$^{1/2}$, as expected for massless Dirac fermions in the

graphene monolayer. The red solid line is a linear fit of the data with Eq. (1), yielding a Fermi velocity of $v_F = (1.14 ± 0.02) \times 10^6$ m/s. Inset: the schematic LL structure of graphene monolayer in the quantum Hall regime.

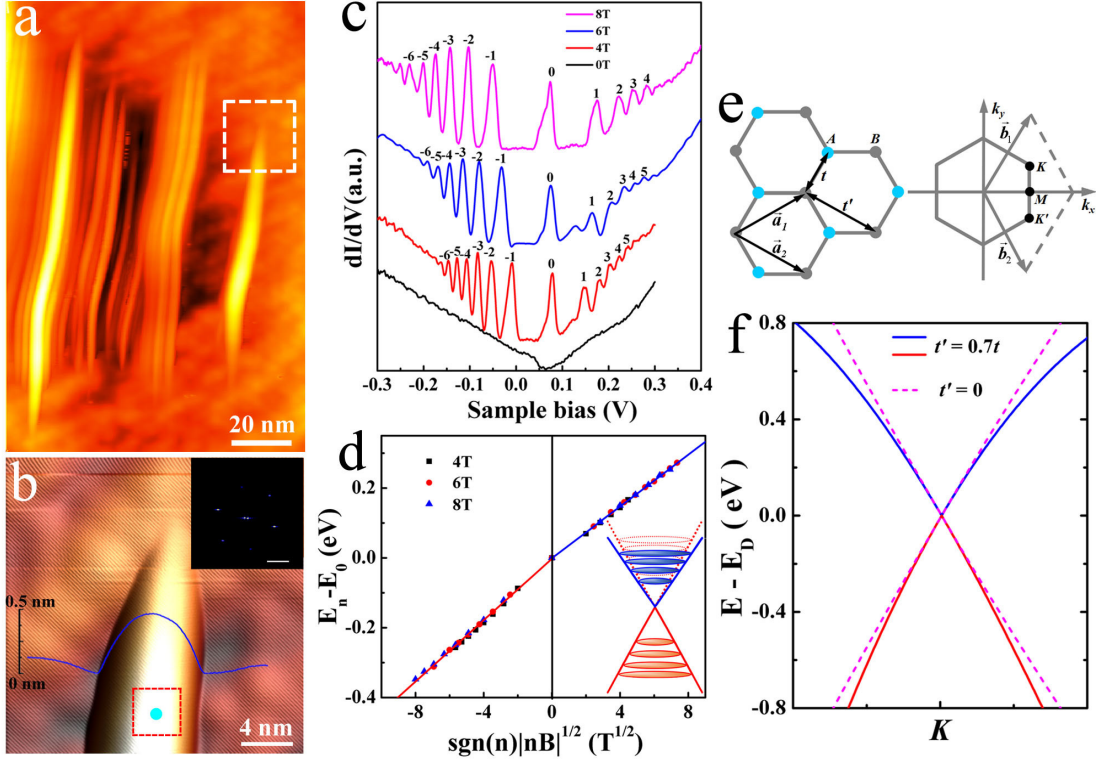

**Figure 2 | STM images and STS spectra of a strained graphene region. a**, A representative STM image of a strained graphene region showing quasi-periodic ripples. **b**, STM image of a strained graphene ripple with large curvature in the red frame in **a** ($V_{sample}$ = 0.57 V and I = 69 pA). The solid blue line is the height profile across the strained graphene region. The inset is the Fast Fourier transform showing the reciprocal-lattice of the graphene lattice in the main panel. The scale bar is 20 $Gm^{-1}$. **c**, STS spectra taken at the light blue solid circle marked in **a** under different magnetic fields. For clarity, the curves are offset in Y-axis and LL indices are marked by the black numbers. **d**, LL peak energies recorded in the light blue dot show

square-root dependence on both the LL index and magnetic field, i.e., sgn($n$)|$nB$|$^{1/2}$, as expected for massless Dirac fermions in graphene monolayer. The red and blue solid line are linear fits of the data with Eq. (1), yielding the Fermi velocities (1.21 ± 0.03) × 10$^6$ m/s and (1.02 ± 0.03) × 10$^6$ m/s, for electrons and holes, respectively. Inset: the schematic LLs structure of the strained graphene sheet in the quantum Hall regime. The $n > 0$ LLs in dotted curves are plotted to show the LL structure with electron-hole symmetry. **e,** Left: lattice structure of graphene, made out of two inequivalent carbon atoms *A* and *B*. $\vec{a}_1$ and $\vec{a}_2$ are the lattice unit vectors. *t* and *t*′ are the nearest-neighbor and next-nearest-neighbor hopping energy, respectively. Right: corresponding Brillouin zone. The Dirac cones are located at the *K* and *K*′ points. $\vec{b}_1$ and $\vec{b}_2$ are the reciprocal-lattice unit vectors. **f,** Electronic dispersion for graphene honeycomb lattice with different *t*′ (*t* = 3 eV). To show the effect of *t*′ clearly, we use an unrealistic value *t*′ = 0.7*t* in panel **f**. Actually, we extract *t*′ = 0.16*t* to account for the electron-hole asymmetry observed in panel **d**.

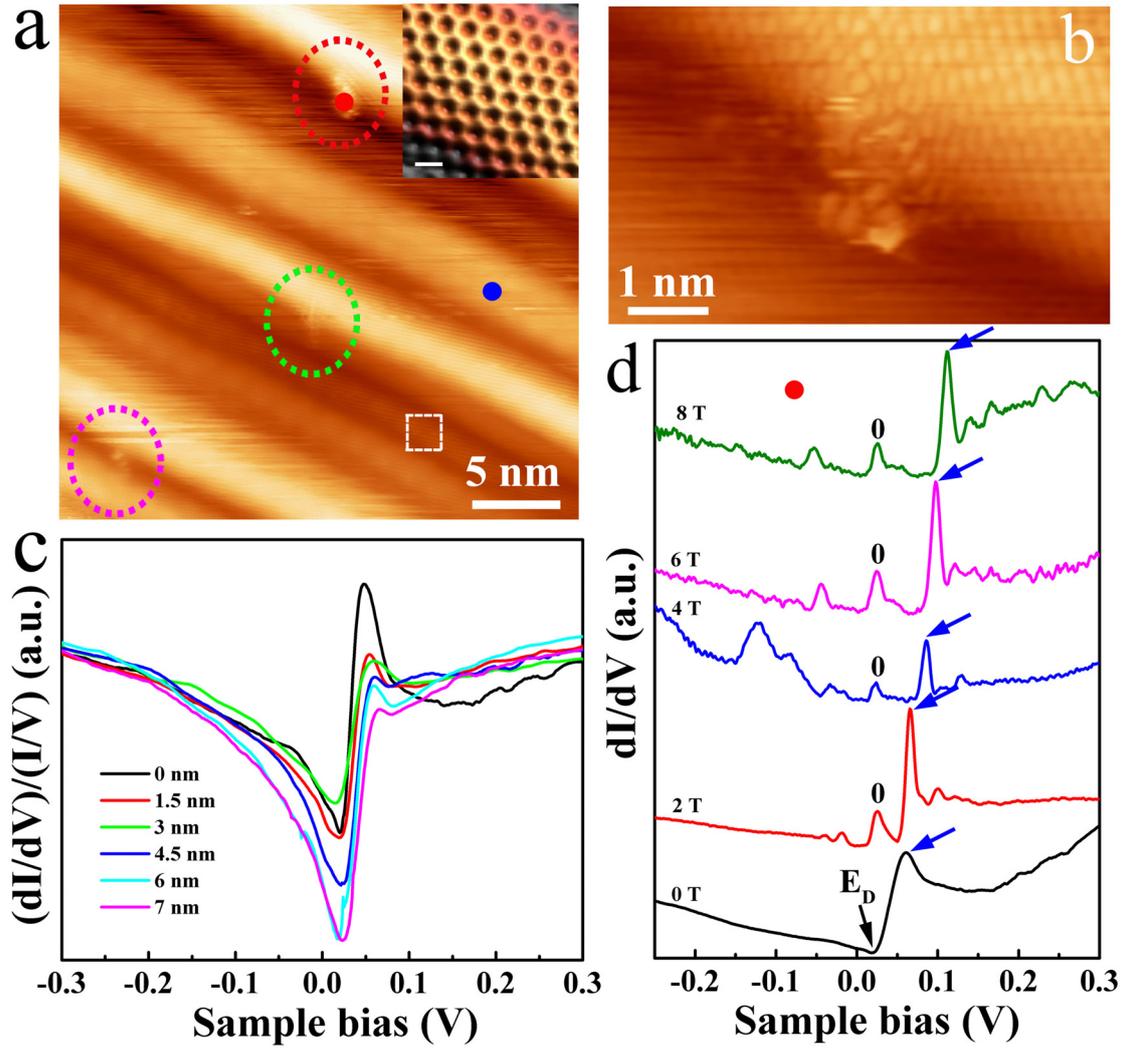

**Figure 3 | STM images and STS spectra of a graphene monolayer with charged defect. a**, A representative STM image of graphene monolayer with three atomic defects ($V_{sample}$ = 0.98 V and $I$ = 290 pA), as marked by the three dashed ellipses. The graphene sheet also exhibits nonuniform nanoscale ripples. The wavelength and amplitude of these nanoscale ripples are shown in Fig. S3. The inset shows an enlarged image in the white frame where we can observe honeycomb lattice and the scale bar is 0.3 nm. **b**, The zoom-in STM image of the defect in the red ellipse in panel **a**. A clear interference pattern, attributing to intervalley scattering, is observed in proximity of the defect. **c**, Normalized STS spectra, (d$I$/d$V$)/($I$/$V$)-$V$, measured on

graphene at different distances from the center of the defect. **d**, STS spectra taken at the red solid dot marked in **a** under different magnetic fields. For clarity, the curves are offset in Y-axis. The blue arrows indicate the resonance peaks of the defect under different magnetic fields. $E_D$ is the Dirac point and the $n = 0$ LL is marked.

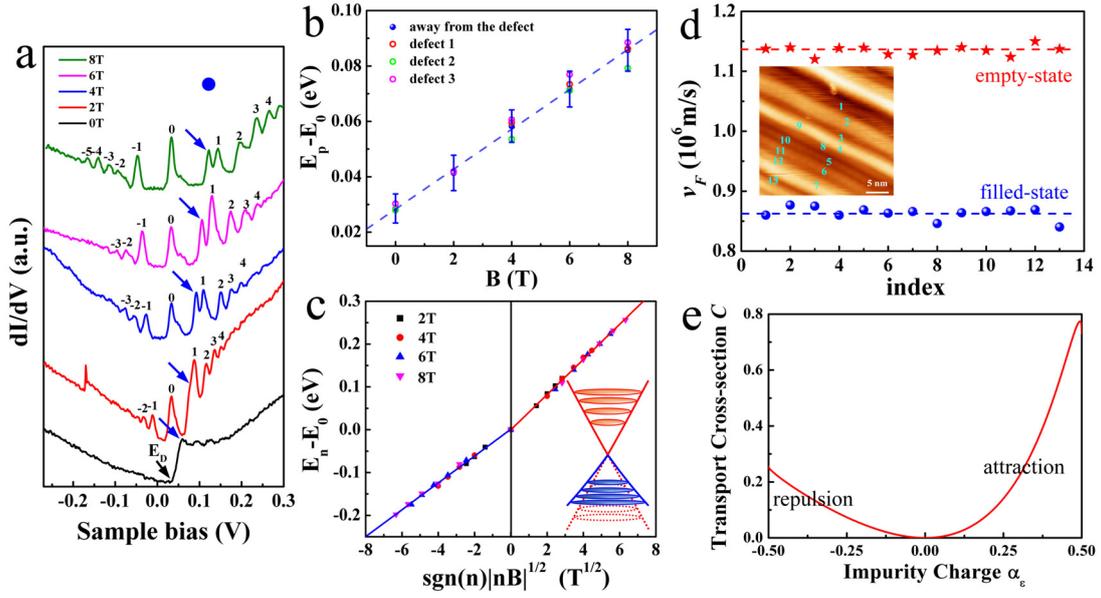

**Figure 4 | STS spectra and electron-hole asymmetry around the charged defects.**

**a**, STS spectra taken under different magnetic fields at the blue solid circle marked in Fig. 3a. LL indices of massless Dirac fermions are marked. The blue arrows indicate the resonance peak of the defect under different magnetic fields. **b** shows the values of $E_p - E_0$ obtained at different positions as a function of the magnetic fields. The red, green, and maganta circles indicate the value of $E_p - E_0$ obtained on the defects within the ellipses of corresponding colors in Fig. 3a. The blue solid line is the guide to eyes. **c**, LL peak energies in panel **a** show square-root dependence on both the LL index and magnetic field, i.e., $\text{sgn}(n)|nB|^{1/2}$. The red and blue solid lines are the linear fits of the data, yielding the Fermi velocities $v_F^h = (1.14 \pm 0.03) \times 10^6$ m/s and $v_F^e = (0.86 \pm$ 0.02) × $10^6$ m/s. Inset: the schematic LL structure of the graphene sheet around the charged-defect in the quantum Hall regime. The $n < 0$ LLs in dotted curves are plotted to show the LL structure with electron-hole symmetry. **d**, This figure summarizes the Fermi velocities of electrons and holes obtained at different positions around the charged defects. The inset shows the STM image with 13 different positions marked by the light blue numbers. **e**, Transport cross-section *C* as a function of the impurity charge $\boldsymbol{\alpha_\varepsilon}$. Impurity charge $\boldsymbol{\alpha_\varepsilon} = \alpha\ \text{sgn}(\varepsilon)$, where $\boldsymbol{\alpha_\varepsilon} > 0$ indicates the attractive interaction between the impurity and the carriers, and $\boldsymbol{\alpha_\varepsilon} < 0$ indicates the repulsive interaction between them.